\documentclass[doublecol,linenumbers]{epl2}
\usepackage{subfigure}
\usepackage{amsmath, amsthm, amssymb}

\title{Intra-layer synchronization in multiplex networks}

\author{L.V. Gambuzza\inst{1} \and M. Frasca\inst{2} \and J. G{\'o}mez-Garde\~{n}es\inst{2,3}}
\shortauthor{L.V. Gambuzza \etal}

\institute{
  \inst{1} Dipartimento di Ingegneria Elettrica, Elettronica e Informatica, University of Catania, Italy\\
  \inst{2} Departamento de F\'{\i}sica de la Materia Condensada,
  University of Zaragoza, Spain\\
  \inst{3} Institute for Biocomputation and Physics of Complex
Systems (BIFI), University of Zaragoza, Spain}

\abstract{We study synchronization of $N$ oscillators indirectly coupled through a medium which is inhomogeneous and has its own dynamics. The system is formalized in terms of a multilayer network, where the top layer is made of disconnected oscillators and the bottom one, modeling the medium, consists of oscillators coupled according to a given topology. The different dynamics of the medium and the top layer is accounted by including a frequency mismatch between them. We show a novel regime of synchronization as intra-layer coherence does not necessarily require inter-layer coherence. This regime appears under mild conditions on the bottom layer: arbitrary topologies may be considered, provided that they support synchronization of the oscillators of the medium. The existence of a density-dependent threshold as in quorum-sensing phenomena is also demonstrated.}

\begin{document}

\maketitle

\section{Introduction}
\label{sec:Intro}
Synchronization is one of the most ubiquitous collective phenomena appearing in natural and artificial systems \cite{boccbook}. Singing crickets, fireflies emitting sequences of light flashes, cardiac pacemakers, circadian rhythms in mammals, firing neurons, chemical systems exhibiting oscillatory variation of the concentration of reagents, applauding audiences, or electrical and electronic devices are all common examples of systems operating in synchrony \cite{SYNC}. In general, all of these examples can be described as systems composed of many units that adjust a particular dynamical property to behave in unison. The interaction among the units is at the core of synchronization since, when isolated, they behave according to their individual rhythms.
In the recent years the way units interact and its influence to the onset of synchrony have been the subject of intense research, where complex networks have been used to account for a variety of interaction patterns \cite{physrepbocca,physreparenas}. These patterns include the modeling of heterogeneity of links, delays in signal interchange, and time-dependent connections.

The main hypothesis underlying the network approach is that the units of a system (modeled as the nodes of a graph) are directly coupled through interactions represented by the network edges \cite{rev:albert,rev:newman}. However, in many physical systems the units interact in an indirect way. For instance, in the Huygens's experiment, historically considered the first report on synchronization \cite{Bennett}, the two pendulum clocks interact through the wooden beam on which they are both mounted. Similarly, communication between cellular populations occurs thanks to small molecules diffused in the medium \cite{mcMillen02}, and chemical oscillators interact through a stirred solution \cite{Taylor09}. Even in the excessive wobbling observed in the opening of the Millennium Bridge in London, the synchronous pacing of the crowd derives from the interaction of the pedestrians with the bridge \cite{millenniumBridge}. Synchrony in this case only occurs for a population density greater than a threshold, a phenomenon which is called as \emph{crowd synchrony}.

Synchronization of indirectly coupled units has been studied in several works. The first evidences of synchronization through indirect coupling were observed in the context of quorum-sensing studies \cite{ojalvo04,demonte07,Taylor09}. For instance, yeast cells, which show a density-dependent transition to synchronous oscillations, only interact by exchanging signaling molecules in the extracellular solution \cite{demonte07}. The studies about the synchronization of periodic oscillators coupled through a common medium 
have been recently extended to
chaotic systems. In this latter case, when two chaotic units are considered, both in-phase and anti-phase synchronization have been numerically \cite{sharma2012,sharma2012AD} and experimentally \cite{sharma2011,suresh2014} observed. When more than two chaotic units are taken into account, phenomena such as phase synchronization, periodic collective behavior and quorum-sensing transition show up \cite{russo}.

In this paper, we consider a multi-layer network benchmark \cite{bocc2014} to model a population of oscillators (described as a network layer) whose units interact indirectly through the coupling to an extended medium (a different layer). In particular, this latter medium is composed of another system of coupled oscillators with a characteristic natural frequency. The main result is that synchronization of the indirectly coupled units (composing the first network layer) is possible even in the absence of coherence with the medium. In addition we show that for synchronization to occur, the presence of an amplitude dynamic variable is needed. As a consequence, synchronization is observed for units with amplitude as well as phase dynamics, but not for ones with phase dynamics only.


\section{The model}
\label{sec:Model}


In most studies on not directly coupled dynamical units a homogeneous distribution of the medium is assumed. While this assumption is reasonable for chemical systems under the hypothesis of well-stirred solutions or biological systems under the hypothesis of fast diffusion of the small molecules, in other contexts (such as genetic oscillators \cite{mcMillen02}) the interactions may be mediated by one agent in the medium for each dynamical unit. Thus, a model in which units are not directly coupled, while the agents in the medium interact, is needed.

\begin{figure}
\centering {\includegraphics[width=7cm]{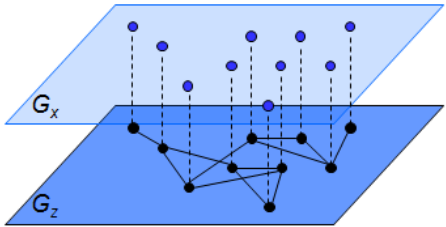}}
\caption{\label{fig:framework2} Representation of the multiplex network consisting of two layers with one-to-one coupling between the layers. In the top layer, called $x$, the nodes only interact with those in the bottom one whereas in the bottom layer $z$ the nodes also interact with other members of the same layer.}
\end{figure}

In this work, we consider this latter scenario and propose a dynamical model incorporating a microscopic description of the interactions of the agents in the medium. In particular, we account for the assumption of inhomogeneous and not passive environment by investigating a system made of two layers, one representing the medium, called layer $z$, and the other, called $x$, the dynamical units. The interaction between layers (medium and units) is as schematically shown in Fig.~\ref{fig:framework2}. Each unit interacts directly with one agent in the medium. Therefore, in terms of the recently developed theory of multi-layer networks \cite{bocc2014,ML1}, our system is termed as a {\em multiplex network} of two layers.

Multiplexes have recently attracted a lot of attention as they are the kind substrates representing better the interaction patterns occurring in many dynamical processes on real complex systems. Examples of the dynamical setups and phenomena studied in multiplex networks include: linear diffusion \cite{ML2,ML5}, congestion and traffic \cite{barthelemy}, evolutionary dynamics \cite{ML3} or epidemics \cite{ML4}.

To extend this knowledge to the realm of synchronization we assume that each node of the multiplex is a Stuart-Landau (SL) oscillator with different natural frequency \cite{RSpre,RSChaos}. In this way the multiplex is formed by $N$ SL units coupled according to the adjacency matrix $A^{z}_{ij}$ in the layer $z$, and $N$ SL oscillators in the layer $x$ that are not directly coupled. Thus, the evolution of the state, $u_{j}^{\alpha}\in \mathbb{C}$, of oscillator $j$ in layer $\alpha(=x,z)$, is given by:

\begin{equation}
\begin{small}
\label{eq:multilayer}
\begin{array}{lll}
\dot{u}^{x}_{j} & = &(a+{\mbox{i}}\cdot\omega_{j}^{x}-|u_{j}^{x}|^2)\cdot u_{j}^{x}+\lambda_{zx}\cdot (u_{j}^{z}-u_{j}^{x})\;,\\
\dot{u}^{z}_{j} & = &(a+{\mbox{i}}\cdot\omega_{j}^{z}-|u_{j}^{z}|^2)\cdot u_{j}^{z}+\lambda_{zx}\cdot (u_{j}^{x}-u_{j}^{z})\\& & +\lambda_z\cdot \sum_{l=1}^{N}A^{z}_{jl}(u_{l}^{z}-u_{j}^{z})\;,
\end{array}
\end{small}
\end{equation}
where $\sqrt{a}$ and $\omega_{j}^{\alpha}$ are respectively the amplitude and the frequency of oscillator $j$ when uncoupled (to account for the frequency mismatch between layer $z$ and $x$, the natural frequency is function of the layer), and $\lambda_z$ and $\lambda_{zx}$ are the coupling between the agents in the medium and the inter-layer coupling respectively.

Let us note that the Stuart-Landau model considered here contains the Kuramoto model \cite{Acebron} (the usual benchmark for the study of synchronization in networks) as a limiting case when the amplitude dynamics vanishes, which occurs when $a$ is large. 

\section{Regime of intra-layer coherence}
\label{sec:existence}

We now investigate the existence of phase synchronization in the multiplex. Our aim is to show that, besides global synchronization, {\em i.e.}, the regime in which all the nodes in the multiplex network are synchronized with each other, a state characterized by intra-layer coherence and inter-layer incoherence is possible. We refer to this regime as intra-layer coherence (ILC), implicitly assuming that there is no coherence between the two layers of the multiplex (otherwise the state of global synchronization is obtained).

The regime showing ILC is rather counterintuitive since it implies that all the oscillators in each layer oscillate in synchrony with a shared frequency $\Omega^{\alpha}$ which, in general, is different from one layer to the other. Moreover, as there are no intra-layer connections in layer $x$, the synchronization of this layer is possible due to the indirect coupling of its units through layer $z$. Therefore, in the ILC regime, the nodes in layer $z$ are mediating for synchronization of layer $x$ nodes, without being synchronized with them.

We first analytically show the existence of ILC regime by rewriting the system (\ref{eq:multilayer}) in polar coordinates ($u_{j}^{\alpha}=\rho_{j}^{\alpha}\exp{{\mbox{i}}\theta_{j}^{\alpha}}$) and focusing on the equations for the phases:

\begin{equation}
\begin{small}
\label{eq:multilayerfasi}
\begin{array}{l}
\dot{\theta}_{j}^{x}=\omega^{x}_j+\lambda_{zx}\frac{\rho_{j}^{z}}{\rho_{j}^{x}}\sin(\theta_{j}^{z}-\theta_{j}^{x})\;,\\
\dot{\theta}_{j}^{z}=\omega^{z}_j+\lambda_{zx}\frac{\rho_{j}^{x}}{\rho_{j}^{z}}\sin(\theta_{j}^{x}-\theta_{j}^{z})+\lambda_z\sum_{l=1}^{N} A^{z}_{jl} \frac{\rho_{l}^{z}}{\rho_{j}^{z}}\sin(\theta_{l}^{z}-\theta_{j}^{z})\;.
\end{array}
\end{small}
\end{equation}
We look for solutions of the type $\theta_{1}^{z}=\theta_{2}^{z}=\ldots=\theta_{N}^{z}$ and $\theta_{1}^{x}=\theta_{2}^{x}=\ldots=\theta_{N}^{x}$, {\em i.e.}, solutions where all the oscillators within each layer are synchronized (this condition includes both the regimes of ILC and global synchronization). Under this hypothesis, we consider two generic nodes $j$ and $l$ in Eqs. (\ref{eq:multilayerfasi}), and we also assume that the frequency of the two nodes are similar to derive:

\begin{equation}
\begin{small}
\label{eq:multilayerfasidiff}
\begin{array}{l}
\frac{d}{dt}(\theta_{j}^{x}-\theta_{l}^{x})=\lambda_{zx}(\frac{\rho_{j}^{z}}{\rho_{j}^{x}}-\frac{\rho_{l}^{z}}{\rho_{l}^{x}})\sin(\theta_{j}^{z}-\theta_{j}^{x})\;,\\
\frac{d}{dt}(\theta_{j}^{z}-\theta_{l}^{z})=\lambda_{zx}(\frac{\rho_{j}^{x}}{\rho_{j}^{z}}-\frac{\rho_{l}^{x}}{\rho_{l}^{z}})\sin(\theta_{j}^{x}-\theta_{j}^{z})\;.
\end{array}
\end{small}
\end{equation}

From these equations we notice that a solution corresponding to global synchronization of the multiplex, $\theta_{1}^{z}=\ldots=\theta_{N}^{z}=\theta_{1}^{x}=\ldots=\theta_{N}^{x}$, is always possible. However, the solution corresponding to  ILC, $\theta_{1}^{z}=\ldots=\theta_{N}^{z}=\theta^{z}$ and $\theta_{1}^{x}=\ldots=\theta_{N}^{x}=\theta^{x}$ with $\theta^{z}-\theta^{x} \neq {\mbox{const.}}$, is only possible provided $\frac{\rho_{1}^{z}}{\rho_{1}^{x}}=\frac{\rho_{2}^{z}}{\rho_{2}^{x}}=\ldots=\frac{\rho_{N}^{z}}{\rho_{N}^{x}}$, {\em i.e.}, the nodes in the same layer must have the same amplitude. Thus, by fixing $\rho_{j}^{z}=\rho^{z}$ $\forall j$ and $\rho_{j}^{x}=\rho^{x}$ $\forall j$, and by looking at the equations of the amplitudes, it is possible to show that the ILC solution cannot be achieved with a stationary amplitude, $\dot\rho^{\alpha}=0$ ($\alpha=x, z$), {\em i.e.}, it cannot be observed in a multiplex composed of Kuramoto oscillators. Under this hypothesis, the equations for the amplitude are:

\begin{equation}
\begin{small}
\label{eq:multilayeramplitude}
\begin{array}{l}
\dot{\rho}^{x}=a {\rho}^{x} - ({\rho}^{x})^3 + \lambda_{zx}\left[{\rho}^{z}\cdot \cos(\theta^{z}-\theta^{x})-{\rho}^{x}\right]\;,\\
\dot{\rho}^{z}=a {\rho}^{z} - ({\rho}^{z})^3 + \lambda_{zx}\left[{\rho}^{x}\cdot \cos(\theta^{x}-\theta^{z})-{\rho}^{z}\right]\;.
\end{array}
\end{small}
\end{equation}
From above it becomes clear that a stationary solution ($\dot{\rho}^{x}=\dot{\rho}^{z}=0$) of Eqs. (\ref{eq:multilayeramplitude}) implies that ${\rho}^{x}={\rho}^{z}$, {\em i.e.}, all the nodes having the same amplitude, and

\begin{equation}
\cos(\theta^{x}-\theta^{z})=\frac{\lambda_{zx}+({\rho}^{x})^2-a}{\lambda_{zx}}\;,
\end{equation}
{\em i.e.}, the difference $\theta^{z}-\theta^{x}$ is constant, contrary to the initial hypothesis. This result points out that the solution cannot be stationary (as it requires time-varying amplitudes) and thus it can only be obtained when the amplitude is a free parameter. This condition is met in Stuart-Landau oscillators, but not in Kuramoto ones.


\section{ILC in multiplex networks}
\label{sec:results}

We now provide numerical evidences of the existence of ILC solutions in a small network and, then, examine the case of larger structures. In all the simulations we assume that the natural frequencies, $\omega_{j}^{\alpha}$, of the nodes are uniformly distributed in $[0.95\cdot \omega^{\alpha}, 1.05\cdot \omega^{\alpha}]$, being $\omega^x=1$ and $\omega^z=2.5$.

Phase synchronization between any pair of oscillators of the multiplex, namely oscillator $j$ of layer $\alpha$ and oscillator $l$ of layer $\beta$, can be measured by the Kuramoto order parameter:

\begin{equation}
r_{jl}^{\alpha\beta}=|\langle e^{{\mbox{i}}\cdot[\theta_{j}^{\alpha}(t)-\theta_{l}^{\beta}(t)]}\rangle_t|\;.
\end{equation}
To get some insight on the behavior of the layers we monitor the intra-layer coherence by defining the Kuramoto order parameter of layer $\alpha$ as:

\begin{equation}
\label{eq:kuramotorh}
r^\alpha=\frac{1}{N(N-1)}\sum_{j,l=1}^N{r_{jl}^{\alpha\alpha}}\;,
\end{equation}
\noindent and the inter-layer coherence as:

\begin{equation}
\label{eq:kuramotorzx}
r^{zx}=\frac{1}{N}\sum_{j=1}^N{r_{jj}^{zx}}\;,
\end{equation}
{\em i.e.}, by averaging the degree of synchronization between all the pairs of nodes connected by the inter-layer links.

\begin{figure}[t!]
\centering {\includegraphics[width=8.75cm]{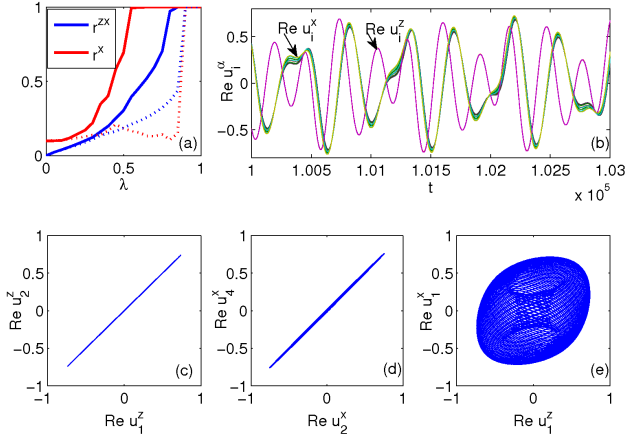}}
\caption{\label{fig:esempiosegnali} (Color online). The ILC regime in a multilayer network with $N=10$ nodes and all-to-all coupling in the bottom layer $z$. (a) Kuramoto order parameters $r^x$ and $r^{zx}$ vs. $\lambda=\lambda_{zx}=\frac{\lambda_z}{5}$. Continuous lines refer to a multilayer network of Stuart-Landau oscillators with $a=1$, whereas dashed ones to purely phase oscillators ($a\rightarrow \infty$), for which ILC does not exist. (b) Waveforms of state variables $\mathrm{Re~} u_{j}^{\alpha}$ for $\lambda=0.7$. (c) Phase plane $\mathrm{Re~} u_{1}^{z}-\mathrm{Re~} u_{2}^{z}$. (d) Phase plane $\mathrm{Re~} u_{2}^{x}-\mathrm{Re~} u_{4}^{x}$. (e) Phase plane $\mathrm{Re~} u_{1}^{z}-\mathrm{Re~} u_{1}^{x}$. In (b)-(e) nodes in each layer are mutually synchronized with the nodes of the same layers, but not with their corresponding counterpart in the other layer.}
\end{figure}

In Fig.~\ref{fig:esempiosegnali} we show the results obtained with a multiplex of $N=10$ units in each layer, where the nodes in layer $z$ are globally connected ($A^{z}_{jl}=1$ $\forall j, l$). The network behavior depends on the coupling coefficients $\lambda_z$ and $\lambda_{zx}$. In this first example the analysis was carried out by simultaneously varying them and keeping their ratio constant, {\em i.e.}, we varied $\lambda$ defined as $\lambda=\lambda_{zx}=\frac{\lambda_z}{5}$ (the more general case of independent coupling coefficients is considered below). 
The Kuramoto order parameters $r^{x}$ and $r^{zx}$ vs. $\lambda$ [see Fig.~\ref{fig:esempiosegnali}(a)] show that $r^x$ grows faster than $r^{zx}$, and consequently there is a range of $\lambda$ values for which the top layer reaches synchronization ($r^x\simeq 1$ ), even if each node of the top layer is not synchronized to its corresponding in the bottom layer ($r^{zx}\ll 1$). When purely phase (Kuramoto) oscillators are considered, this latter regime is not observed as the curves of $r^x$ and $r^{zx}$ are similar [dashed lines in Fig.~\ref{fig:esempiosegnali}.(a)]. The waveforms obtained for $\lambda=0.7$ [see Fig.~\ref{fig:esempiosegnali}(b)] confirm that the ILC regime is only attainable together with non-stationary amplitudes. Fig.~\ref{fig:esempiosegnali}(b) also shows that the nodes in each layer are synchronized with a frequency different from one layer to the other. Intra-layer synchronization without inter-layer coherence is also clear from the phase planes of Figs.~\ref{fig:esempiosegnali}(c)-(e).

We now consider a larger multiplex network with layers of $N=100$ nodes and systematically vary the two parameters $\lambda_z$ and $\lambda_{zx}$. We monitor the difference between the Kuramoto order parameters in Eqs. (\ref{eq:kuramotorh}) and (\ref{eq:kuramotorzx}), {\em i.e.}, $\Delta r= r^{x}-r^{zx}$, as a function of $\lambda_z$ and $\lambda_{zx}$. Large values of $\Delta r$ indicate the appearance of the ILC regime in a region of the parameter space. In fact, also referring to the example reported in Fig. \ref{fig:esempiosegnali}(a), $\Delta r$ is large if $r^{x}>r^{zx}$, that is, if the synchronization level within layer $x$, $r^{x}$, is greater than the inter-layer coherence measured by $r^{zx}$. On the contrary, $\Delta r$ is small if the two measures $r^{x}$ and $r^{zx}$ have similar values, pointing out that the multiplex is either desynchronized or globally synchronized.

\begin{figure}[t!]
\centering
\includegraphics[width=4.2cm]{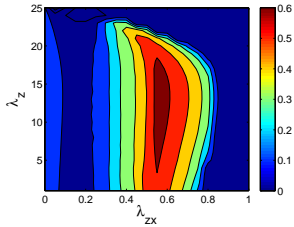}
\caption{\label{fig:AllToAll} (Color online). ILC in a multiplex network with $N=100$ and all-to-all topology in the bottom layer $z$. The contour plot shows the bifurcation diagram of $\Delta r$ vs. $\lambda_z$ and $\lambda_{zx}$.}
\end{figure}

The behavior of $\Delta r$ is reported in Fig. \ref{fig:AllToAll} which shows the appearance of the ILC regime for values of $\lambda_{zx}$ laying approximately between $0.4$ and $0.75$ and for a large range of $\lambda_z$ values. In the whole region of parameters studied, the oscillators in layer $z$ are synchronized with each other. Instead, the oscillators in layer $x$ are not synchronized for $\lambda_{zx} < 0.4$, synchronized with each other but not with their corresponding units in layer $z$ approximately in the range $0.4 < \lambda_{zx} < 0.75$ and $\lambda_z< 20$, and globally synchronized with those of layer $z$ (that is, all the oscillators, independently from the layer to which they belong, run at the same shared frequency), otherwise.

The regime of ILC is elicited by the modulation of the state variables of the bottom layer by the units of the top layer. Each oscillator in layer $z$ is influenced by two terms, one representing the coupling with the corresponding top layer oscillator and one the coupling from its layer $z$ neighbors. These two terms are competing as the two layers have different natural frequencies. An equilibrium between these two terms leads to ILC as oscillators in layer $z$ should allow the passage of information needed to synchronize those in layer $x$ without synchronizing with them. On the contrary, when one of these forces is too strong or too weak, ILC is not possible.

\begin{figure}[t!]
\centering
\subfigure[]{\includegraphics[width=4.2cm]{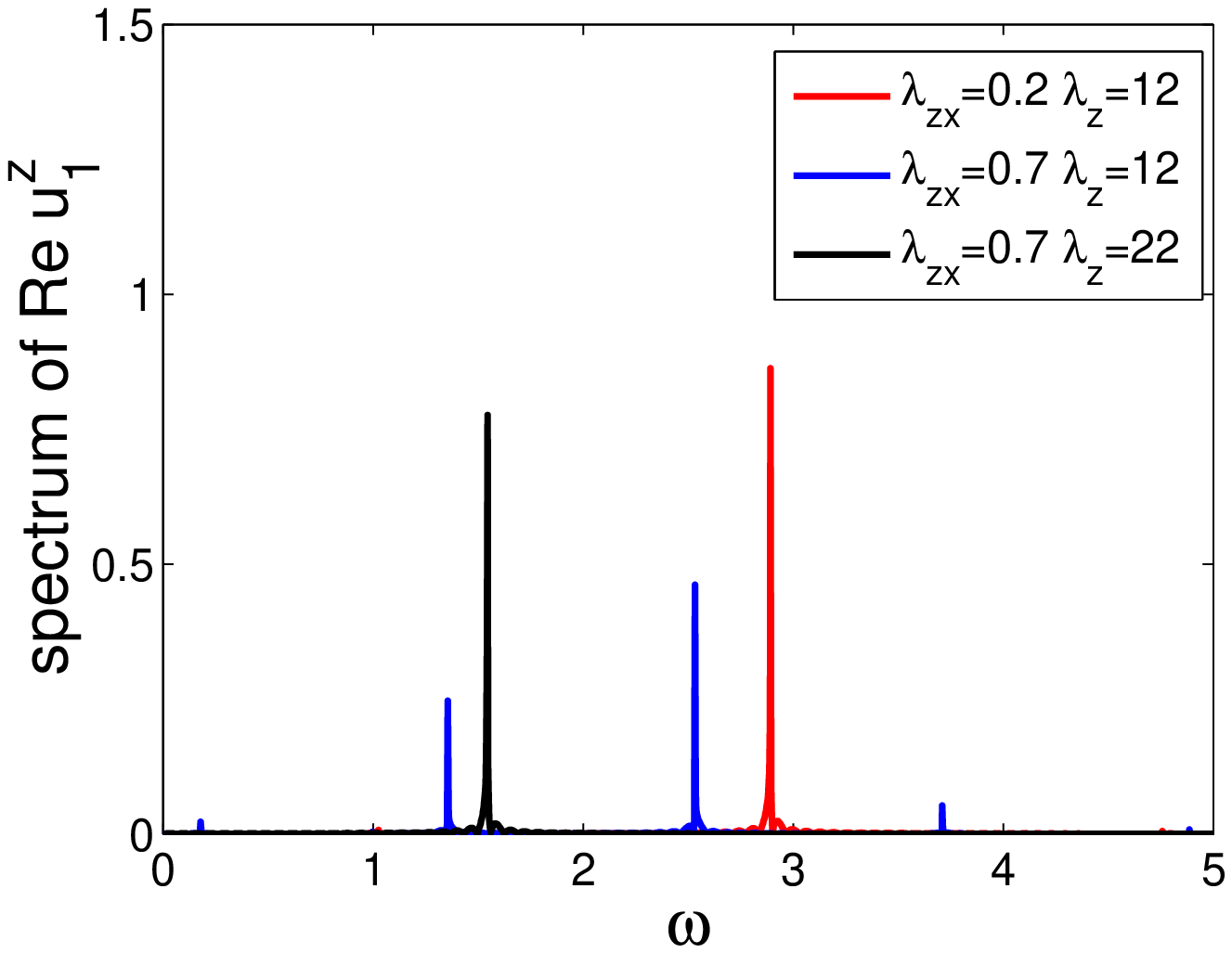}}
\subfigure[]{\includegraphics[width=4.2cm]{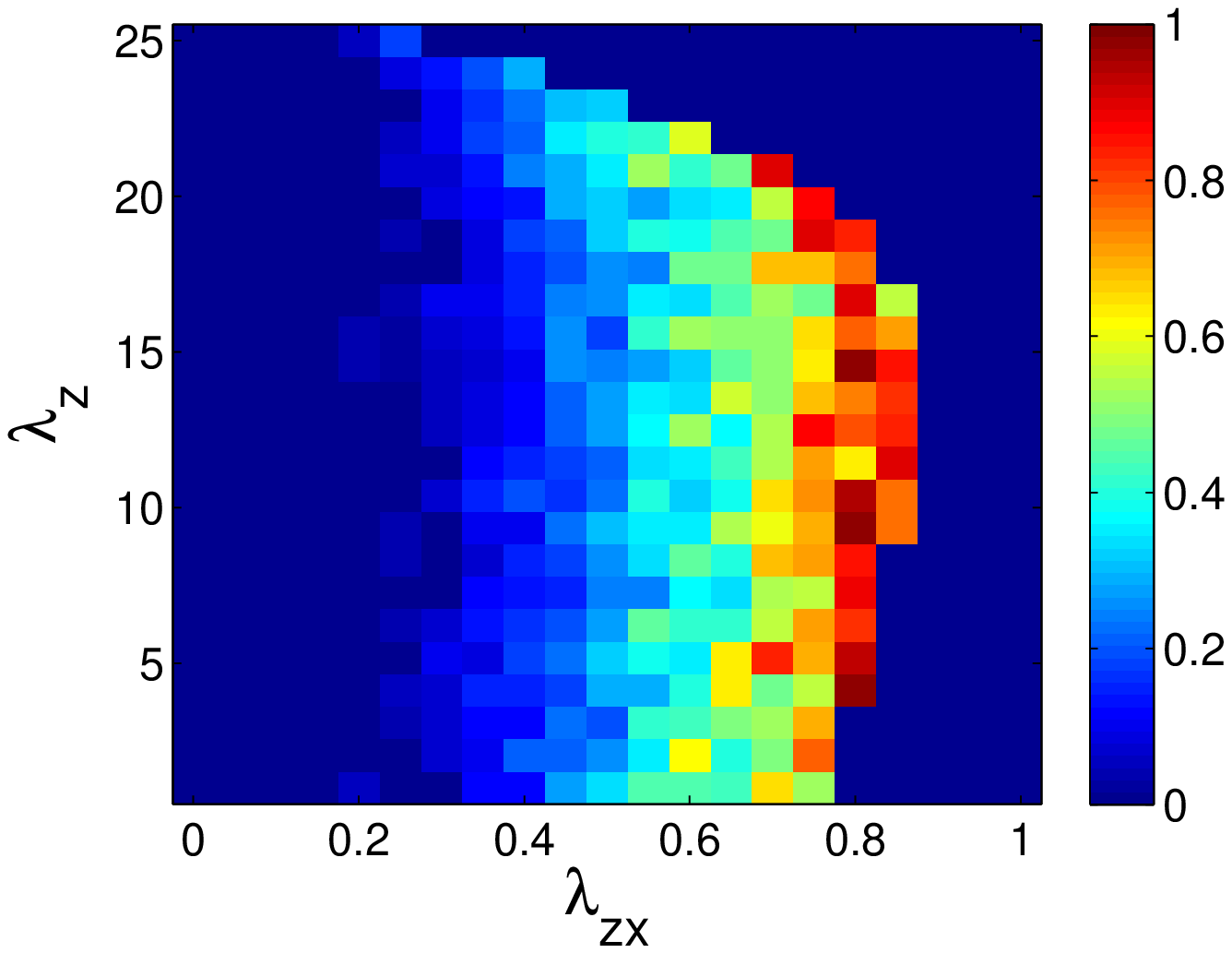}}
\caption{\label{fig:spectra} (Color online). (a) Spectrum of $\mathrm{Re~} u_{1}^{z}$ at several pairs of $\lambda_z$ and $\lambda_{zx}$, $\lambda_z=12$ and $\lambda_{zx}=0.2$ (red line), $\lambda_z=12$ and $\lambda_{zx}=0.7$ (blue line), $\lambda_z=22$ and $\lambda_{zx}=0.7$ (black line). (b) Ratio between the second and the first largest components of the spectra of $\mathrm{Re~} u_{1}^{z}$, $S_2^z/S_1^z$, as a function of  $\lambda_{z}$ and  $\lambda_{zx}$.
}
\end{figure}

To gain a deeper knowledge on the roots of the ILC regime we now focus on the evolution of frequencies of the oscillators in each layer. In particular, we focus on the analysis of the frequency spectrum of the oscillators in the bottom layer obtained from the time evolution of $\mathrm{Re~} u_{j}^{z}$. Fig.~\ref{fig:spectra}(a) reports the spectrum of $\mathrm{Re~} u_{1}^{z}$ (given the all-to-all structure of the bottom layer, the choice of the oscillator to analyze is totally arbitrary) for several pairs of values of $\lambda_z$ and $\lambda_{zx}$. When $\lambda_{z}=12$ and $\lambda_{zx}=0.7$ (blue line), ILC occurs and the spectrum is characterized by two significant components, one near the natural frequency of layer $z$, i.e., $\omega^z$, and one near that of layer $x$, {\em i.e.}, $\omega^x$. This clearly shows that the presence of a significant component at the frequency of the top layer is associated to the appearance of ILC. In the other cases reported the spectrum has only one significant component. In particular, if the inter-layer coupling is too weak, i.e., $\lambda_{zx}=0.2$ and $\lambda_z=12$ (red line), this component is close to $\omega^z$ since the top layer is not significantly influencing the bottom layer. For $\lambda_{zx}=0.7$ and $\lambda_z=22$ (black line) the multiplex network is globally synchronized and the common frequency shared by the two layers lays between the two values $\omega^z$ and $\omega^x$.

When the inter-layer and intra-layer interactions are both strong, the tendency to synchronization of each pair of connected oscillators in the two layers is strong and thus the system tends towards global synchronization without showing ILC. Therefore, the onset of ILC can be revealed by the analysis of the spectrum. More in details, we indicate as $S_1^z$ the amplitude of the largest peak, in the spectrum of $\mathrm{Re~} u_{1}^{z}$ and as $S_2^z$ the second largest peak and monitor the ratio $S_2^z/S_1^z$ as a function of  $\lambda_{z}$ and  $\lambda_{zx}$. The result is shown in Fig.~\ref{fig:spectra}(b), where, as expected, the ILC regime corresponds to high values of the ratio $S_2^z/S_1^z$ and the border between ILC and global synchronization is signaled by the largest values of the former ratio.

\section{Complex networked media}

We now consider different types of interaction topologies for the medium (layer $z$) and, in particular, investigate the onset of the ILC regime in (unweighted and undirected) scale-free (SF) and Erd\"os-R\'enyi (ER) networks. To this aim, we make use of a network model that interpolates between these two types of networks \cite{JesusModelA}. The model is based on the combination of uniform and preferential linking and allows to obtain a one-parameter family of networks. Depending on the model parameter $p$, the networks generated change from a power-law degree distribution ($p=0$) to a Poissonian one ($p=1$) in a continuous way. As the main structural properties such as the average shortest path length and the second moment do not vary linearly with $p$, we focus on some significant values of $p$ (not equally spaced). 

\begin{figure}[t!]
\centering
\subfigure[]{\includegraphics[width=4.25cm]{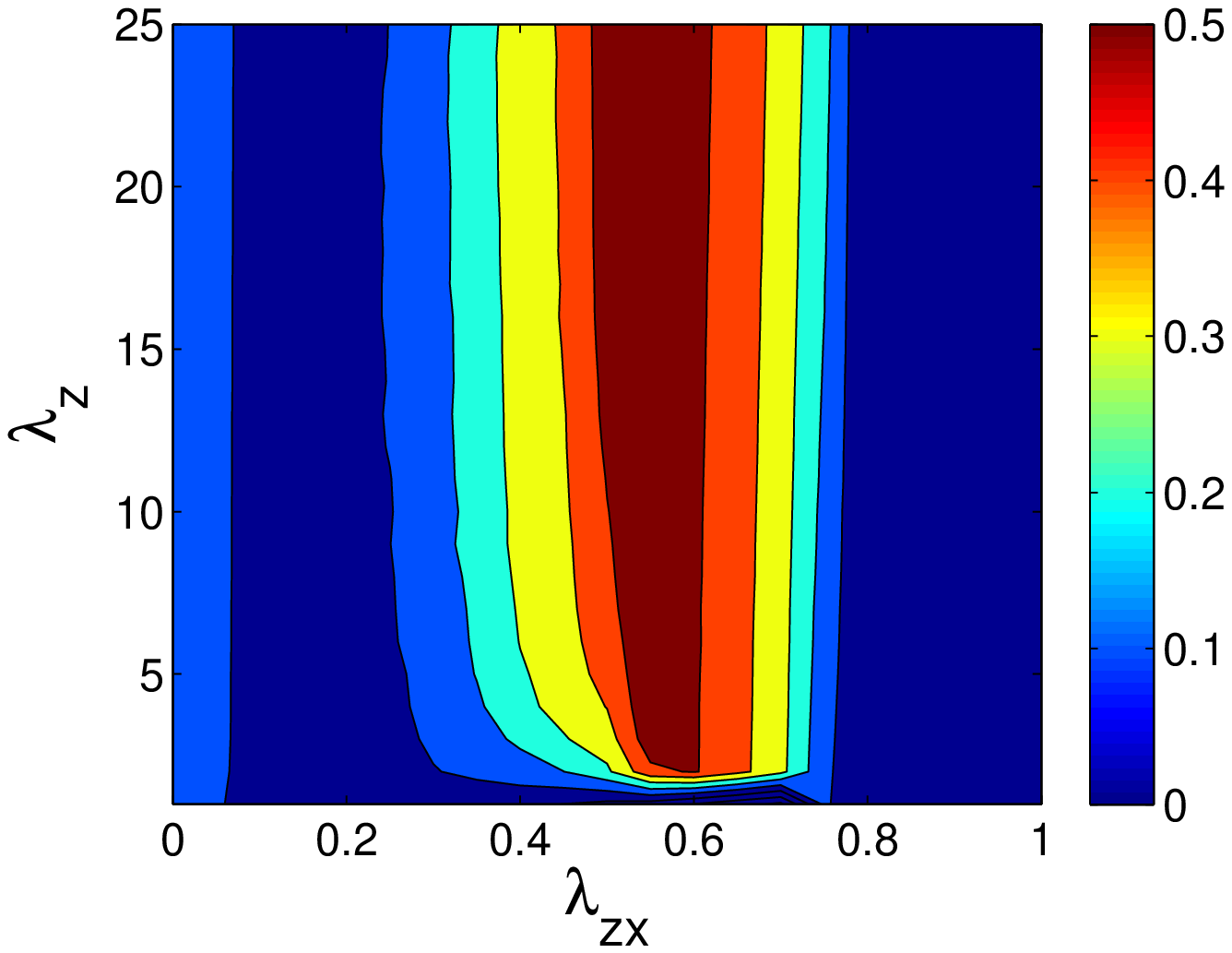}}
\subfigure[]{\includegraphics[width=4.25cm]{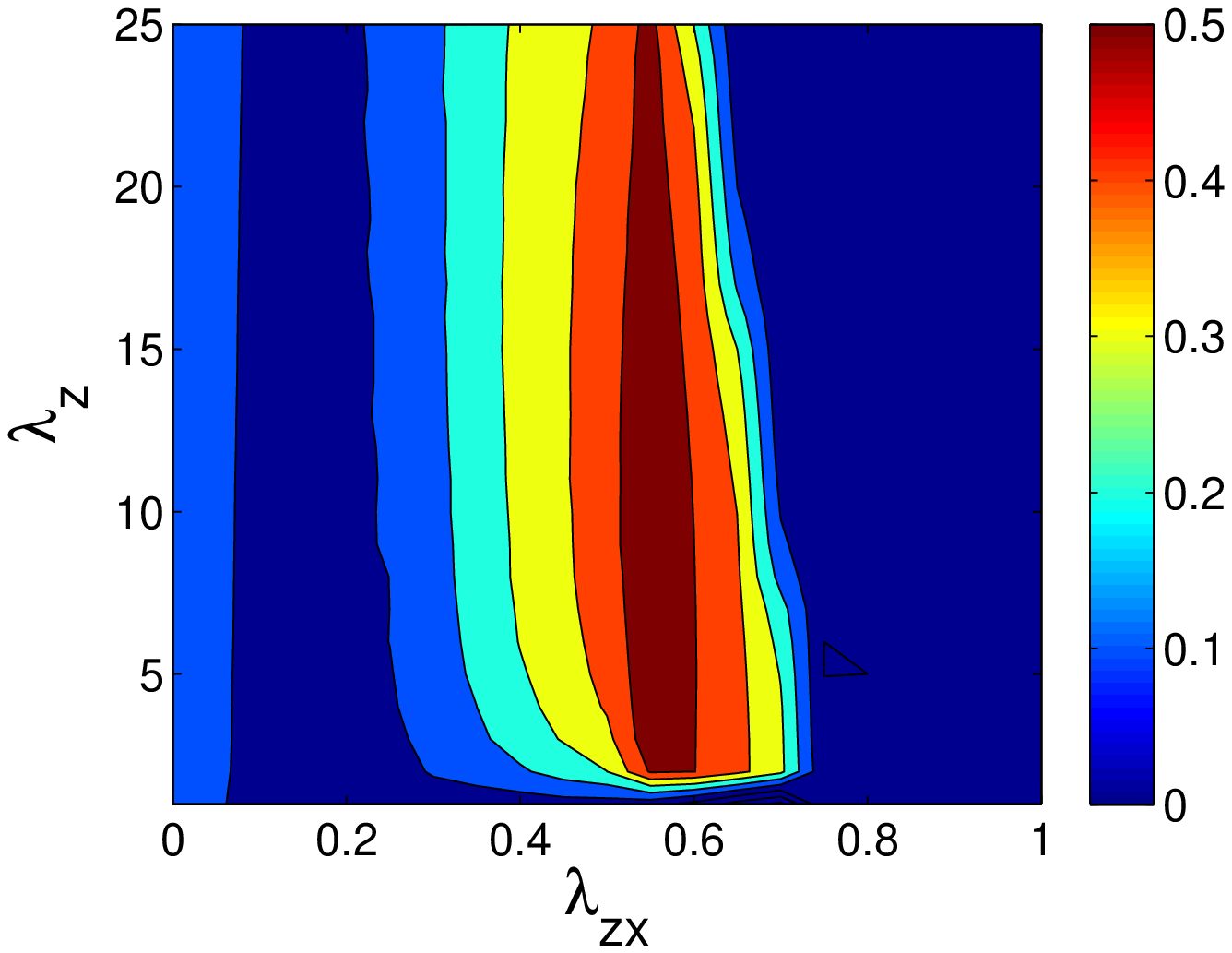}}
\subfigure[]{\includegraphics[width=4.25cm]{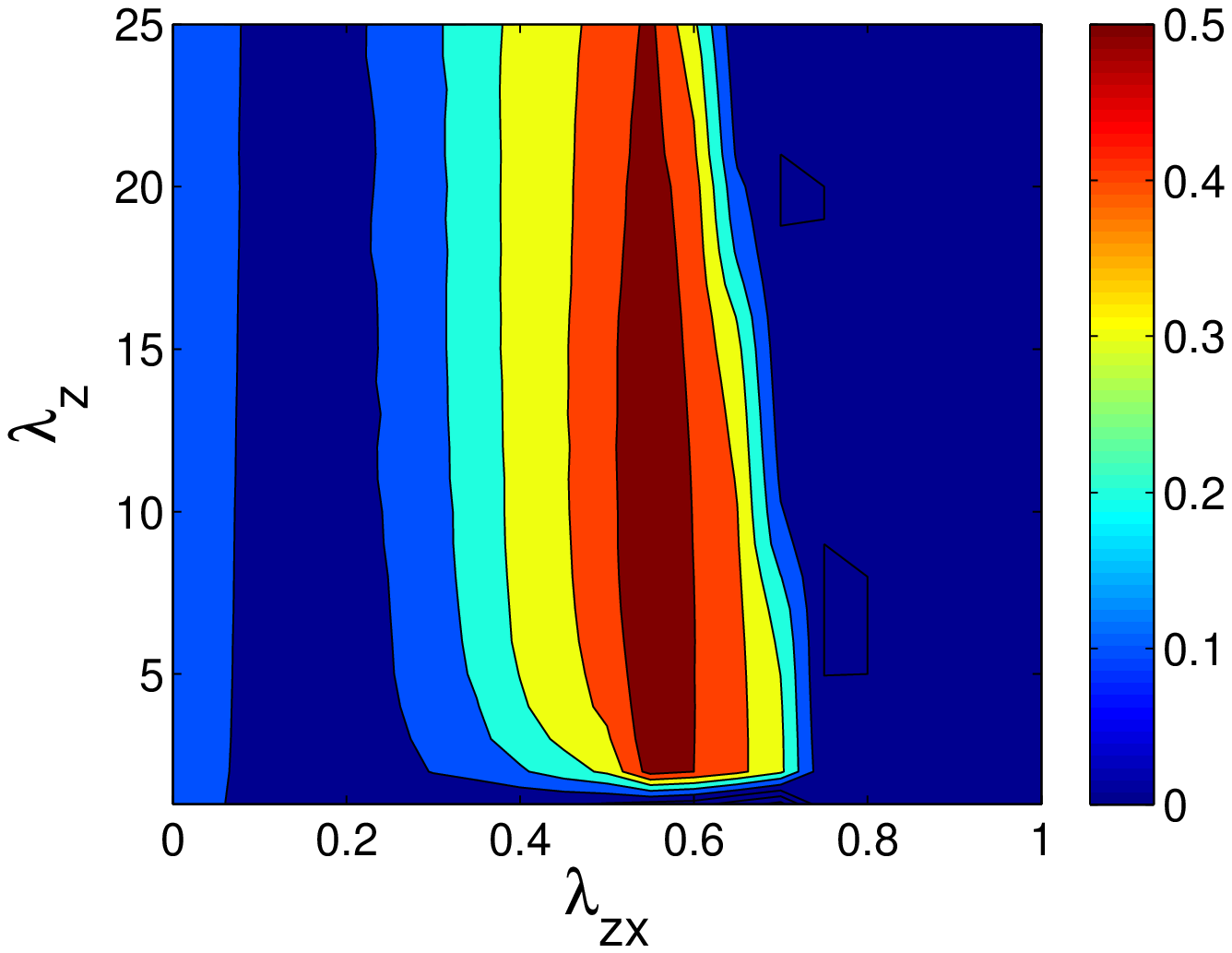}}
\subfigure[]{\includegraphics[width=4.25cm]{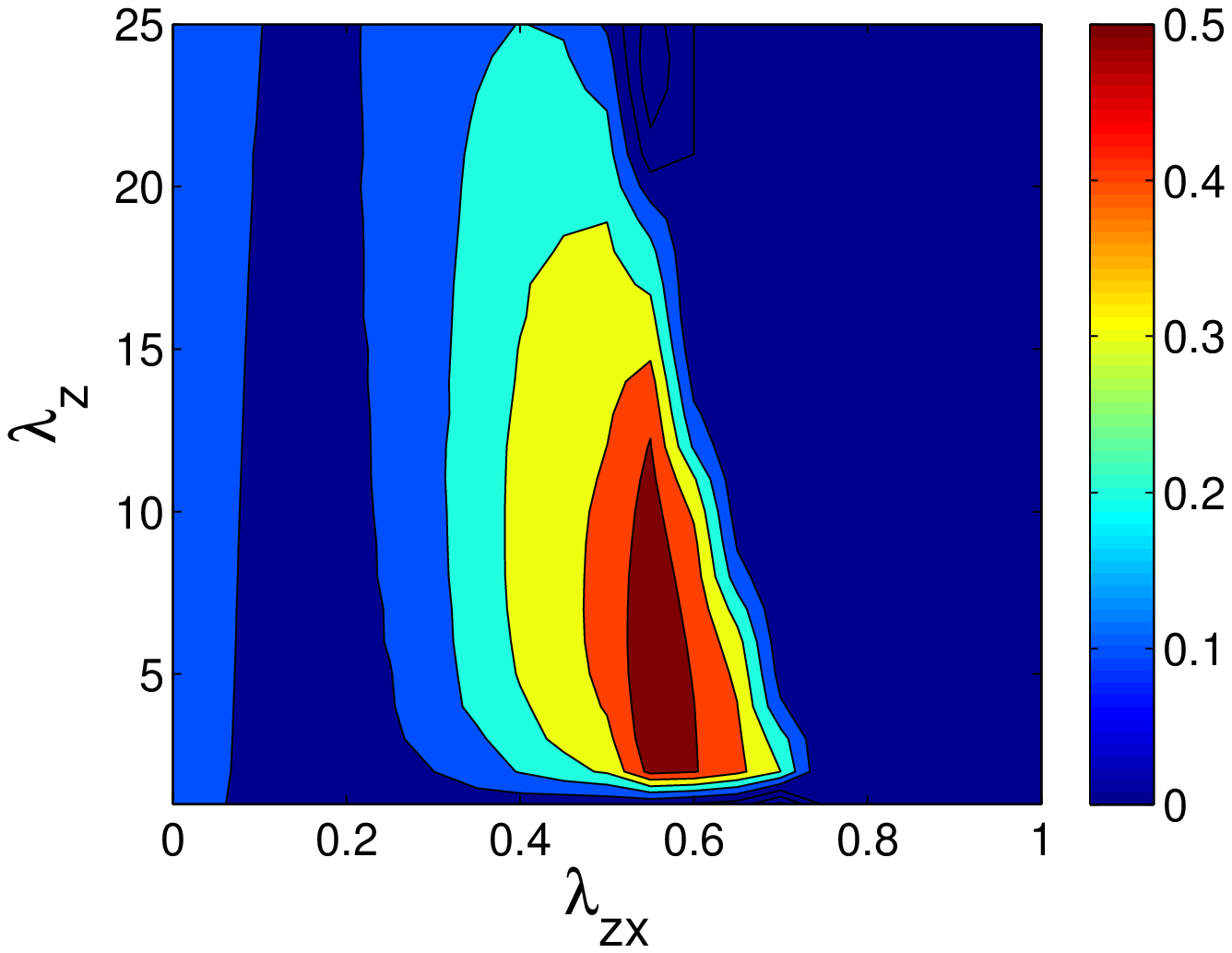}}
\caption{\label{fig:confrontoReti} (Color online). Bifurcation diagram of $\Delta r$ vs. $\lambda_z$ and $\lambda_{zx}$ for multiplex networks with $N=100$ and $\langle k \rangle=8$. The bottom layer $z$ is generated by model A with: (a) $p=0$ (SF network); (b) $p=0.3$; (c); $p=0.6$; (d) $p=1$ (ER network).}
\end{figure}

As it can be observed in Fig. \ref{fig:confrontoReti} the ILC regime appears in all the cases. The region of the parameter space characterized by ILC decreases continuously when going from SF to ER networks. This result indicates that the heterogeneity of the degree distributions, controlled through the interpolation parameter $p$, is the responsible for the different area and that by increasing the heterogeneity of the network we favor the onset of ILC. This finding points out the generality of the ILC regime in multiplex networks and the role of heterogeneity in favoring the onset of it. Let us note that we have checked that degree-degree correlations seem to have no effect on the onset of the ILC regime. In fact, when an uncorrelated SF network, obtained by randomly rewiring the edges of the network used in Fig. \ref{fig:confrontoReti}(a) while preserving the degree distribution, is analyzed, the same behavior (not shown) is obtained.

Finally, given the biological/chemical examples in which the model proposed applies, we have studied the influence that the density of agents in the media has on the onset of ILC. Our aim is to find a density-dependent threshold in a similar fashion to those quorum sensing-like transitions to synchronization, typically induced by the indirect coupling provided by the medium. In fact, many systems of units indirectly coupled through a medium experience a transition from a state to another as a function of the number of individuals or their density. For instance, in bacteria, the response to a stimulus is correlated to population density, a mechanism acting as a decision-making process and called quorum sensing ~\cite{Taylor09}. A similar mechanism, {\em i.e.}, crowd synchrony, is also found in the transition from the quiescent (or disordered) state to the synchronous oscillatory state in populations of oscillators, where the consequence of an increase of the number of units is a larger coupling between them. As increasing intra-layer coupling favors the onset of ILC, we expect that a similar density-dependent transition also occurs in our multiplex.

\begin{figure}[h]
\centering
\subfigure[]{\includegraphics[width=4.25cm]{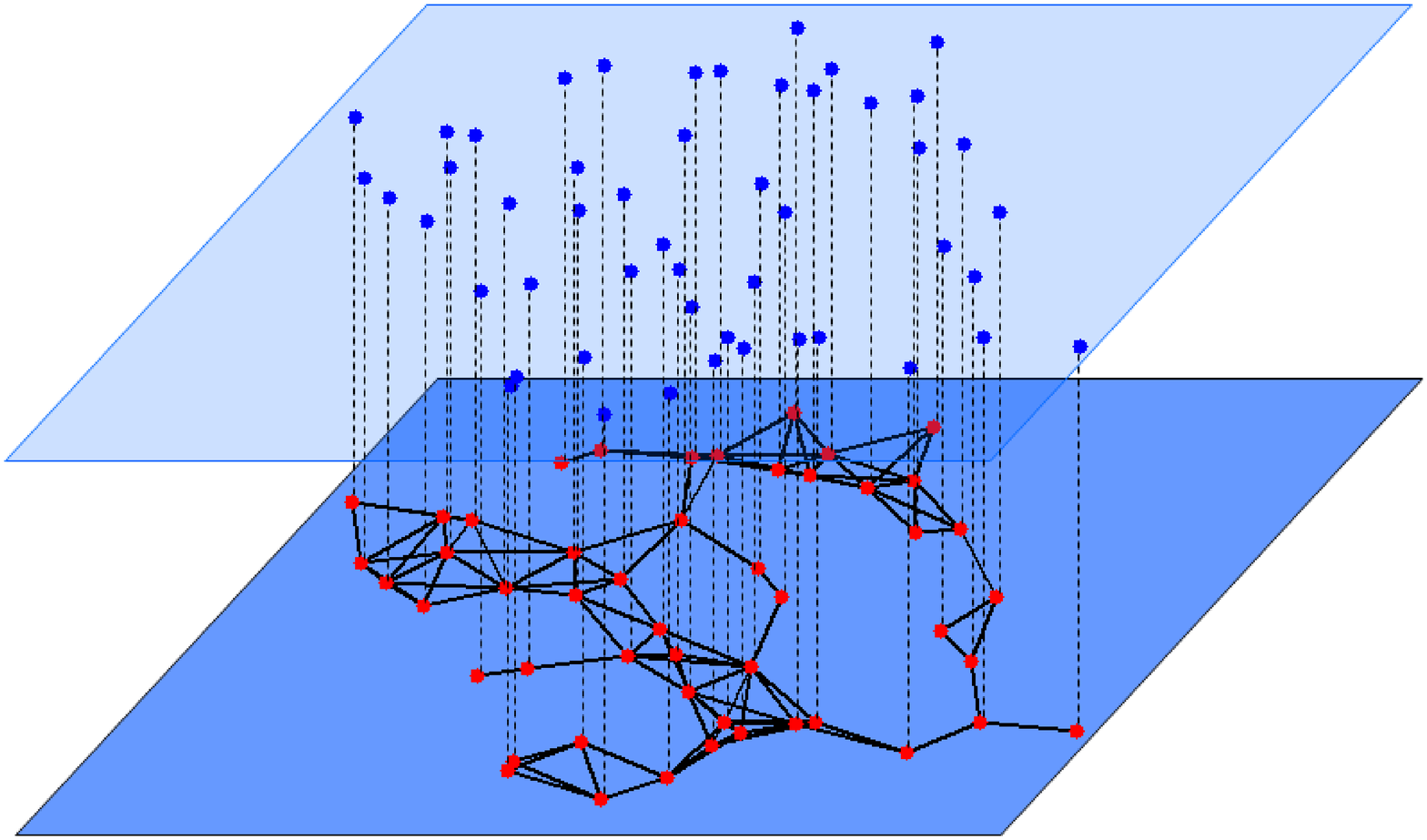}}
\subfigure[]{\includegraphics[width=4.25cm]{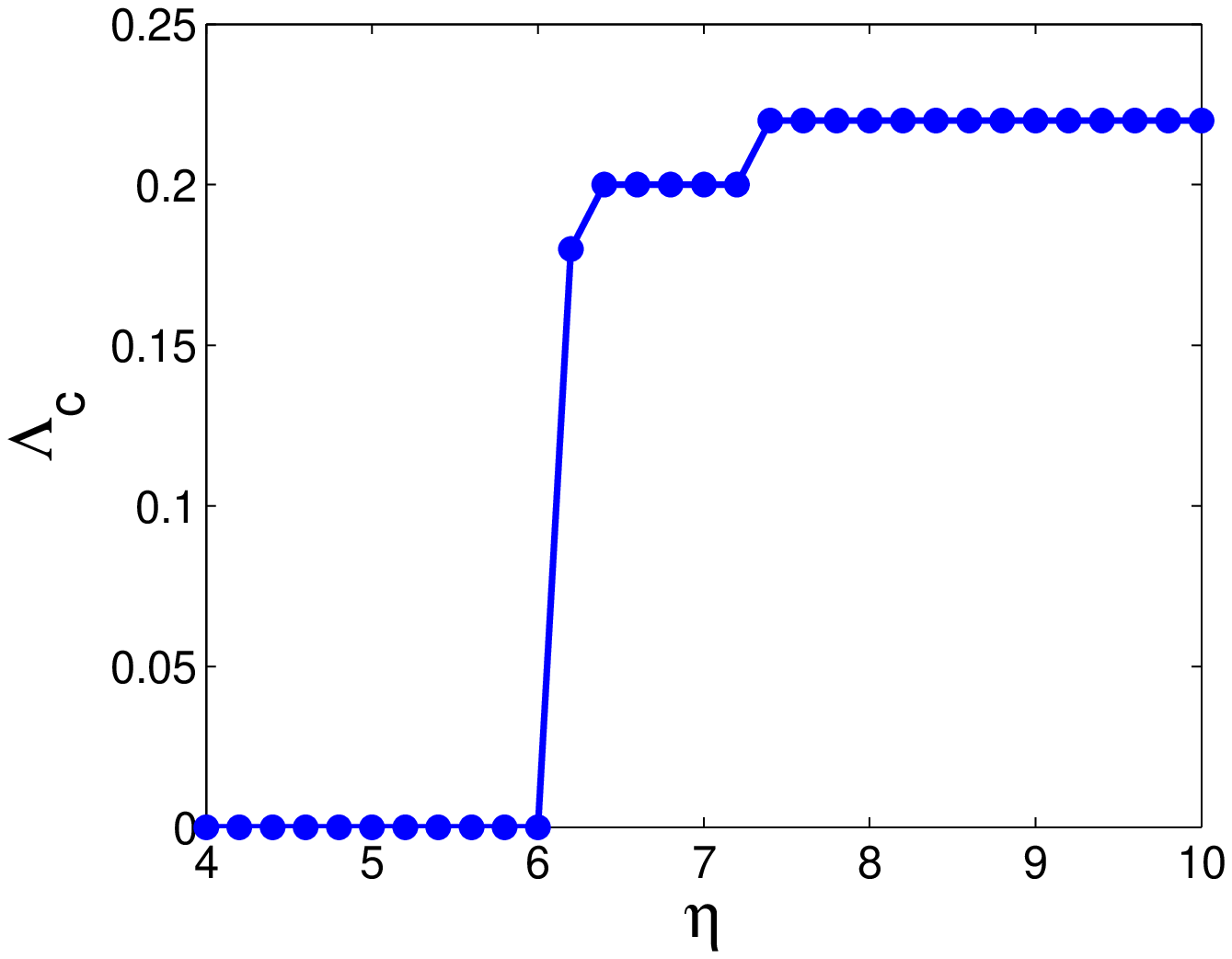}}
\caption{\label{fig:deltaLambda} (Color online). (a) An example of a multilayer network where the bottom layer is a random geometric graph. For the sake of visualization a network with only $N=50$ nodes is displayed. (b) Behavior of $\Lambda_c$ vs. density $\eta$ for a network with $N=100$ and $\lambda_z=2$.}
\end{figure}

We have considered a multiplex network in which the topology of layer $z$ is defined by a Random Geometric Graph \cite{RGG}, {\em i.e.}, a spatial graph in which the nodes are randomly distributed in a planar space of size $L\times L$ with a density given by $\eta=\frac{N}{L^2}$ and each pair of nodes is connected only if their Euclidean distance is less or equal than a given threshold $r$ [see Fig.~\ref{fig:deltaLambda}(a)].
In our analysis, we have kept constant the number of nodes, $N$, and varied the density $\eta$ by changing $L$. To study the onset of a fully developed regime of ILC as function of the density of the particles in the medium, $\eta$, we have run simulations at a fixed value of $\lambda_z$ while varying $\lambda_{zx}$.  The degree of intra-layer synchronization is monitored as described below.

Starting from the typical scenario of ILC shown in Fig.~\ref{fig:esempiosegnali}(a), we observe that $r^x$ reaches values close to $1$ before than $r^{zx}$. Thus, a measure indicating the existence of ILC is given by a large difference in the values of $\lambda_{zx}$ for which $r^x$ and $r^{zx}$ approach $1$. We have thus defined $\lambda_{zx}^1$ as $\lambda_{zx}^1=\min \{ \lambda_{zx}:r_{zx}(\lambda_{zx})>0.95 \}$ and $\lambda_{zx}^2$ as $\lambda_{zx}^2=\min \{ \lambda_{zx}:r_{x}(\lambda_{zx})>0.95 \}$, and monitored the difference between these two values, indicated as $\Lambda_c=\lambda_{zx}^2-\lambda_{zx}^1$.

In Fig.~\ref{fig:deltaLambda}(b) we show the trend of $\Lambda_c$ as a function of the density $\eta$. The plot clearly shows the existence of a density-dependent threshold, $\eta_c$, for the appearance of ILC. In particular, below the value $\eta_c\simeq 6.6$ no ILC regime is observed, whereas above this threshold ILC develops after a very sharp transition. Thus, as expected, the increase of the density of units in the medium clearly acts as a promotor of the ILC state. Note, however, that for $\eta>\eta_c$ the value of $\Lambda_c$ remains roughly constant. This indicates that the width of the region of $\lambda_{zx}$ values in which ILC occurs, for the value of $\lambda_z$ at work [$\lambda_z=2$ in Fig.~\ref{fig:deltaLambda}(b)], do not change with $\eta$ once $\eta>\eta_c$.

\section{Conclusions}
\label{sec:conc}

Summarizing, we have analyzed synchronization in a population of $N$ oscillators indirectly coupled through an inhomogeneous medium. In particular, the medium has its own dynamics, which is of the same type of the oscillators (periodic, when uncoupled), but with a different natural oscillation frequency. The system has been modeled as a multiplex network formed by two layers with the same number, $N$, of nodes, so that each node of a layer is connected to its counterpart in the other layer. We have shown the onset of intra-layer synchronization without inter-layer coherence, {\em i.e.} a state in which the nodes of a layer are synchronized between them without being synchronized with those of the other layer.

Intra-layer synchronization is shown to be a unique dynamical state appearing in systems which can be modeled as multiplex networks. Its robustness has been tested as this regime is commonly observed independently from the topology of the layer corresponding to the medium, provided that it supports synchronization, although the exact region in the parameter space in which it appears depends on its structural features.
The results hold even in the case of sparse connectivity patterns for the top layer, not able to guarantee the synchronization without the action of the bottom layer. The most interesting result showing intra-layer synchronization is that in which the top layer is a collection of $N$ unconnected oscillators. We have shown that intra-layer synchronization is also possible in this scenario, thus obtaining the synchronization of the collection of isolated nodes without synchronizing with the bottom layer.

We have also shown that the presence of an amplitude dynamics, allowing synchronization of units not directly connected, is fundamental, as the regime of intra-layer synchrony is not observed in purely phase oscillators, such as those in the Kuramoto model. This fact, together with its intrinsic relation with multiplex networks, makes intra-layer synchronization states particularly relevant for their unique features in the context of synchronization studies in complex networks.

\medskip

\acknowledgments
This work has been partially supported by the Spanish MINECO under projects FIS2011-25167 and FIS2012-38266-C02-01; and by the European FET projects PLEXMATH (Ref. 317614) and MULTIPLEX (Ref. 317532). J.G.-G. is supported by the Spanish MINECO through the Ram\'on y Cajal Program.

\end{document}